\documentclass[a4paper,11pt]{article}
\usepackage{pos}
\usepackage{graphicx}
\usepackage{amsfonts, amsmath, amsthm}
\usepackage{slashed}
\usepackage{euscript}
\usepackage[font=small,labelfont=bf,figurename=Figure]{caption}
\usepackage[english]{babel}

\title{Elastic neutrino scattering on nuclear systems as a probe of neutrino electromagnetic interactions}

\author[a]{Konstantin Kouzakov}
\author*[b]{Fedor Lazarev}
\author[b]{Alexander Studenikin}

\affiliation[a]{Department of Nuclear Physics and Quantum Theory of Collisions, Faculty of Physics, Lomonosov Moscow State University,\\
Moscow 119991, Russia}
\affiliation[b]{Department of Theoretical Physics, Faculty of Physics, Lomonosov Moscow State University,\\
Moscow 119991, Russia}

\emailAdd{kouzakov@srd.sinp.msu.ru}
\emailAdd{lazarev.fm15@physics.msu.ru}
\emailAdd{studenik@srd.sinp.msu.ru}

\abstract{We study the electromagnetic contribution to elastic neutrino-proton and neutrino-neutron scattering processes. The neutrino electromagnetic charge, magnetic, electric, and anapole form factors of both diagonal and transition types in the mass basis are taken into account in the present formalism. When treating the nucleon electromagnetic vertex, we take into account not only the charge and magnetic form factors of a nucleon, but also its electric and anapole form factors. We examine how the effects of the neutrino magnetic moment can be disentangled from those of the strange quark contributions to the nucleon’s weak neutral current form factors.}

\FullConference{%
  41st International Conference on High Energy physics - ICHEP2022\\
  6-13 July, 2022\\
  Bologna, Italy
}

\begin{document}
\maketitle

\section{Introduction}
There are a large number of experiments investigating both neutrino oscillations and their interactions. In both cases, it is important to theoretically investigate neutrino scattering on various targets \cite{nue,nup}, since scattering processes are either a tool for detecting neutrino fluxes: the processes of neutrino scattering on a nucleon studied in this work contribute to the signals of such experiments as MiniBooNE \cite{MiniBooNE} and registration of supernova neutrinos in JUNO \cite{JUNOYB}; or a tool for studying fundamental interactions of neutrinos: in this work, the contribution of the electromagnetic properties of neutrinos is studied. The latter neutrino properties emerge in different extensions of the Standard Model, and they include millicharges and charge radii, electric, magnetic and anapole moments~\cite{Window}.

\section{Cross sections} 
We consider the process where an ultrarelativistic neutrino with energy $E_\nu$ originates from a source (reactor, accelerator, the Sun, etc.) and elastically scatters on a nucleon in a detector at energy-momentum transfer $q=(T,{\bf q})$. If the neutrino is born in the source in the flavor state $|\nu_\ell\rangle$, then its state in the detector is $|\nu_\ell(\mathcal{L})\rangle=\sum_{k=1}^3U^*_{\ell k}\exp(-i\frac{m_k^2}{2E_\nu}\mathcal{L})|\nu_k\rangle,$ where $\mathcal{L}$ is the source-detector distance. We assume the target nucleon to be free and at rest in the lab frame. The matrix element of the transition $\nu_\ell(\mathcal{L})+N\to\nu_j+N$, where $N$ is either a proton or a neutron, due to weak interaction is given by
%
\begin{eqnarray}
\label{M_weak}
     \mathcal{M}_j^{(w)}&=&\frac{G_F}{\sqrt{2}}U^*_{\ell j}e^{-i\frac{m_j^2}{2E_\nu}\mathcal{L}}
     \bar{u}^{(\nu)}_{j,\lambda'}(k')\gamma^\mu(1-\gamma^5)u^{(\nu)}_{j,\lambda}(k)J_\mu^{\rm (NC)}.
\end{eqnarray}
%
Here $J_\lambda^{\rm (NC)}$ is a weak neutral current of a nucleon, $\bar{u}^{(\nu)}_{j,\lambda'}(k')=u^{(\nu)\dag}_{j,\lambda'}(k')\gamma^0$, where $u^{(\nu)}_{j,\lambda}(k)$  is the bispinor amplitude of the massive neutrino state $|\nu_j\rangle$ with 4-momentum $k$ and spin state $\lambda$.
The matrix element due to electromagnetic interaction is
%
\begin{eqnarray}
\label{M_el-m}
     \mathcal{M}_j^{(\gamma)}=-\frac{4\pi\alpha}{q^2}\sum_{k=1}^3U^*_{\ell k}e^{-i\frac{m_k^2}{2E_\nu}\mathcal{L}}\bar{u}^{(\nu)}_{j,\lambda'}(k')\Lambda^{({\rm EM};\nu)\mu}_{jk}(q)u^{(\nu)}_{k,\lambda}(k)J_\mu^{(\rm EM)},
\end{eqnarray}
%
where $J_\mu^{(\rm EM)}$ is the electromagnetic current of the nucleon and $\Lambda^{({\rm EM};\nu)\mu}_{jk}(q)$ is the neutrino electromagnetic vertex. The nucleon currents can be expanded as follows:
%
\begin{equation}
\label{currents1}
     \begin{aligned}
     J^{(\rm NC)}_\lambda(q)=\bar{u}^{(N)}_{s'}(p')\Lambda^{({\rm NC};N)}_\lambda(-q)u^{(N)}_{s}(p,s),\qquad
     J^{(\rm EM)}_\lambda(q)=\bar{u}^{(N)}_{s'}(p')\Lambda^{({\rm EM};N)}_\lambda(-q)
     u^{(N)}_{s}(p,s),
     \end{aligned}
\end{equation}
%
where $\Lambda^{({\rm NC};N)}_\lambda(q)$ and $\Lambda^{({\rm EM};N)}_\lambda(q)$ are nucleon neutral weak and electromagnetic vertexes, respectively. We consider the following neutrino and nucleon vertexes~\cite{Window,Nowakowski,Alberico09}:
%
\begin{equation}
    \begin{aligned}
    \Lambda^{({\rm EM};\nu)fi}_\mu(q)=&(\gamma_\mu-q_\mu\slashed{q}/q^2)[f^{fi}_Q(q^2)+f^{fi}_A(q^2)q^2\gamma_5]-i\sigma_{\mu\nu}q^\nu[f^{fi}_M(q^2)+if^{fi}_E(q^2)\gamma_5],\\
    \Lambda^{({\rm EM};N)}_\mu(q)=&\gamma_\mu F_Q^N(q^2)-\frac{i}{2m_N}\,\sigma_{\mu\nu}q^\nu F_M^N(q^2)\\
    &+\frac{1}{2m_N}\,\sigma_{\mu\nu}q^\nu\gamma_5 F_E^N(q^2)-\left(q^2\gamma_\mu-q_\mu\slashed{q}\right)\gamma_5 \frac{F_A^N(q^2)}{(2m_N)^2},\\
    \Lambda^{({\rm NC};N)}_\mu(q)=&\gamma_\mu F_1^N(q^2)-\frac{i}{2m_N}\,\sigma_{\mu\nu}q^\nu F_2^N(q^2)-\gamma_\mu\gamma_5 G_A^N(q^2)+\frac{1}{m_N}\,G_P^N(q^2)q_\mu\gamma_5,
    \end{aligned}
\end{equation}
%
where $m_N$ is the nucleon mass, $f^{fi}_{Q,M,E,A}$ ($F_{Q,M,E,A}^N(q^2)$) are the neutrino (nucleon) charge, magnetic, electric and anapole form factors, respectively, $F_{1,2}^N(q^2)$ and $G_{A,P}^N$ are, respectively, the Pauli, Dirac, axial and pseudoscalar weak neutral-current nucleon form factors.

When evaluating the cross section, we neglect the neutrino masses. Since the final massive state of the neutrino is not resolved in the detector, the differential cross section measured in the scattering experiment is given by
\begin{eqnarray}
\label{cr_sec}
     \frac{d\sigma}{dT}=\frac{\left|\mathcal{M}\right|^2}{32\pi E_\nu^2m_N}, \qquad
%
%
     \left|\mathcal{M}\right|^2=\sum_{j=1}^3\left|{\mathcal{M}}_j^{(w)}+{\mathcal{M}}_j^{(\gamma)}\right|^2,
\end{eqnarray}
where averaging over initial and summing over final spin polarizations is assumed.
The expressions for the differential cross section will be given elsewhere \cite{PhysANucl}. 
\section{Numerical illustration}
Below we present numerical calculations for elastic neutrino-nucleon scattering. We restrict ourselves to the case of electromagnetic charge and magnetic form factors of a nucleon, accounting for the relation between the nucleon weak neutral-current and electromagnetic form factors
%
\begin{equation}
    \begin{aligned}
    F_{1,2}^{p(n)}(q^2)&=\left(\frac{1}{2}-2\sin^2\theta_W\right)F_{Q,M}^{p(n)}(q^2)-\frac{1}{2}F_{Q,M}^{n(p)}(q^2)-\frac{1}{2}F_{1,2}^S(q^2),\\
    G_{A}^{p(n)}(q^2)&=\pm\frac{1}{2}G_{A}
    (q^2)-\frac{1}{2}G_{A}^S(q^2),
    \end{aligned}
\end{equation}
%
where $F_{1,2}^S$, $G_{A}^S$ are strange form factors of the nucleon, and $G_A$ is the nucleon axial form factor for the charged-current neutrino-nucleon scattering. We use the parametrization that can be found in Ref.~\cite{Kosmas16} and references therein.

\begin{center}
	\begin{minipage}[h]{0.49\linewidth}
		\center{\includegraphics[width=1\linewidth]{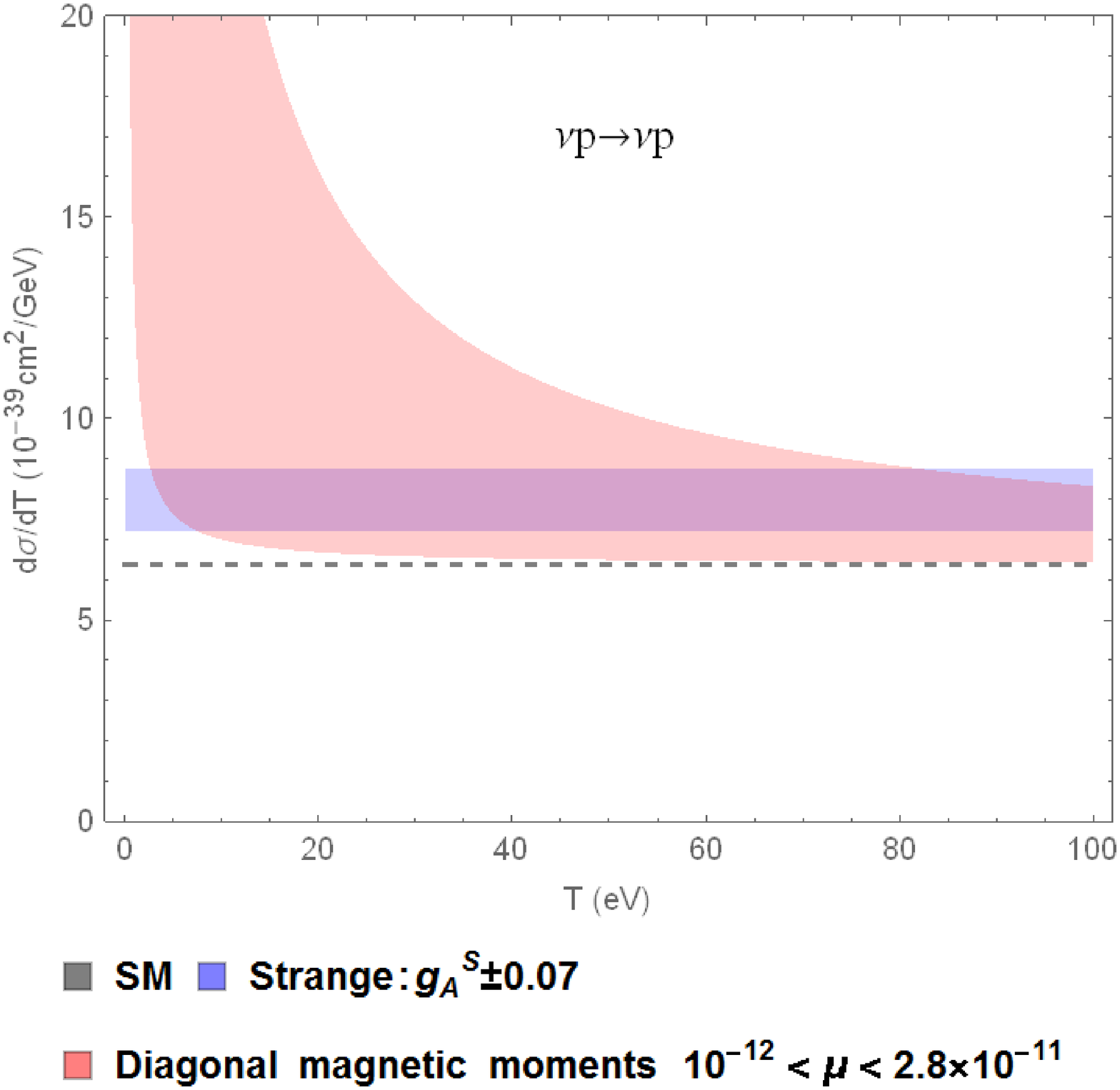} \\ (a)}
	\end{minipage}
	\hfill
	\begin{minipage}[h]{0.49\linewidth}
		\center{\includegraphics[width=1\linewidth]{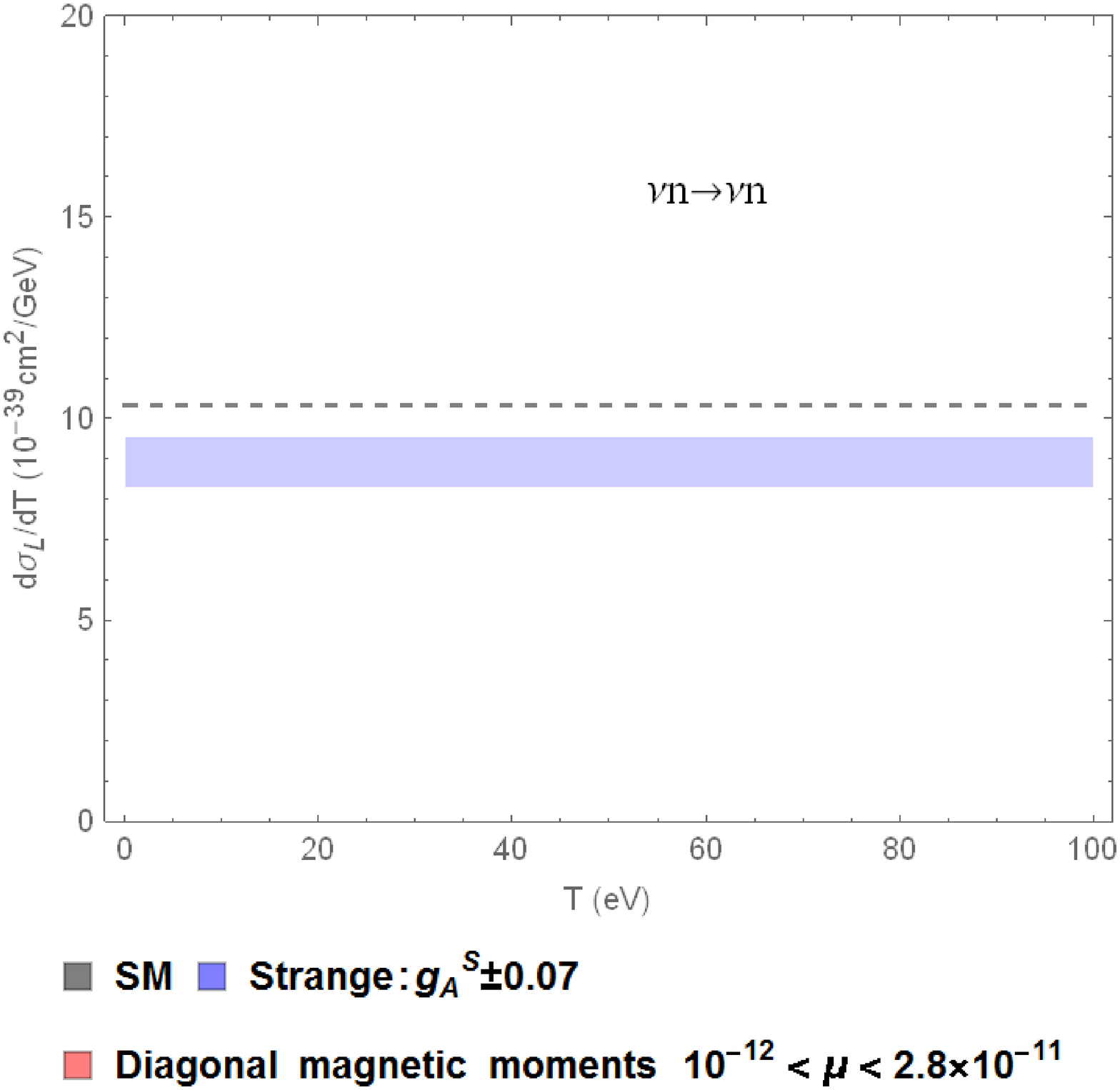} \\ (b)}
	\end{minipage}
	\captionof{figure}{The differential cross sections of neutrino scattering on (a) a proton and (b) a neutron at small  energy transfer values, accounting for diagonal neutrino magnetic moments.}
	\label{MM}
\end{center}
In Fig. \ref{MM}, we show the numerical results in the case of a zero source-detector distance, with and without contribution from both the strange form factors of the nucleon and the neutrino magnetic moment in the range $10^{-12}\mu_B<\mu_{\nu}<2.8\times10^{-11}\mu_B$ that reflects the upper astrophysical and laboratory bounds on the $\mu_\nu$ value~\cite{PDG2022}. As anticipated, the neutrino magnetic moment practically does not have an effect in the case of neutrino-neutron scattering. It can be seen that in the case of the neutrino-proton scattering the effect of the neutrino magnetic moment can be misinterpreted as that of the strange quark contribution to the proton weak neutral-current form factors and vice versa. Therefore it is important to study the energy dependence of the differential cross section in a rather wide range of the proton recoil energies in order to disentangle these effects in experiment.

%
\section{Summary}
The  neutrino electromagnetic properties have been accounted for in the differential cross sections of elastic neutrino-nucleon scattering. On this basis, the numerical results have been obtained for the cases of a nonzero neutrino magnetic moment and a nonzero $s$-quark contribution to the nucleon form factors. 

The results of our work can be used in the studies of neutrino interactions and oscillations in matter, detection of supernova neutrinos using elastic neutrino scattering on protons~\cite{JUNOYB}, the study of the anapole moment of the nucleon, the search for the electric dipole moment of the neutron, the search for the electromagnetic characteristics of neutrinos. 

%
%
\section*{Acknowledgments}
The work is supported by the Russian Science Foundation under grant No.22-22-00384.
F. Lazarev acknowledges the support from the National Centre for Physics and Mathematics (Sarov, Russia).

\end{document}